Research Article

# AI in Work-Based Learning: Understanding the Purposes and Effects of Intelligent Tools Among Student Interns

John Paul P. Miranda[1*], Rhiziel P. Manalese[1], Sheila M. Geronimo[2],
Vernon Grace M. Maniago[1], Charlie K. Padilla[1], Aileen P. De Leon[1],
Santa L. Merle[1], Mark Anthony A. Castro[1]

1. **Pampanga State University**, Pampanga, Philippines
2. **University of the East**, Manila, Philippines

\* **Correspondence:**
John Paul P. Miranda, Pampanga State University, jppmiranda@pampangastateu.edu.ph

## ABSTRACT

This study examined how student interns in Philippine higher education use intelligent tools during their OJT. Data were collected from 384 respondents using a structured questionnaire that asked about AI tool usage, task-specific applications, and perceptions of confidence, ethics, and support. Analysis of task-based usage identified four main purposes: productivity and report writing, communication and content drafting, technical assistance and code support, and independent task completion. ChatGPT was the most commonly used AI tool, followed by Quillbot, Canva AI, and Grammarly. Students reported moderate confidence in using AI and applied these tools selectively and ethically during OJT tasks. This indicate that AI tools assist student interns in various OJT activities related to work-readiness. The study suggests that higher education programs include AI literacy and onboarding. Clear policies and fair access to AI tools are important to support responsible use and prepare students for future careers.

*Keywords: AI in education, generative AI, AI tools, OJT, internship, practicum, immersion, Philippines*

# INTRODUCTION

On-the-job training (OJT) is a required part of many college programs in the Philippines. It gives students the chance to apply what they have learned in school to real work situations (Hingpit, 2024; Loc et al., 2025). During OJT, student interns are asked to do a range of tasks such as writing reports (Debliquy et al., 2025), solving problems (Wang et al., 2024), communicating with staff or clients (Castillo-Núñez et al., 2024; Schartel Dunn & Lane, 2019), and providing technical support (Catacutan & Tuliao, 2020; Keawtavon et al., 2023). These experiences help them build work-related skills and understand how companies or offices operate (Yamauchi et al., 2023).

While many students use artificial intelligence (AI) tools in academic settings (Vieriu & Petrea, 2025), it is not yet clear how frequently or in what ways these tools are used during their OJT. Various AI-powered applications are now commonly used to support writing, editing, and design tasks in schoolwork (Al-Sofi, 2024; Bringula, 2023; Garcia, n.d.; Marzuki et al., 2023). However, the use of such tools in completing real workplace tasks assigned during internships remains underexplored (Hernandez et al., 2025a).

Several studies have examined the integration of AI in specialized internship programs. For example, (Öncü et al., 2025) described how AI is used to simulate real-life medical scenarios through virtual standardized patients like ChatGPT-4o, which helps interns practice clinical decision-making and crisis management without risk to real patients. Similarly, AI-nurse systems assist in preparing medical students for emergency situations by simulating overnight on-call duties (Weaver et al., 2025). In the field of engineering, AI tools have been adopted to accelerate the learning process and enhance the effectiveness of practical training sessions (Salutina et al., 2024). Another study used AI to provide real-time feedback and align internship tasks with labor market demands, although it did not directly investigate how student interns themselves engage with AI tools (Abas et al., 2025). The emergence of virtual internships in some universities also suggests new possibilities for AI integration during internship experiences (Sanahuja Vélez et al., 2017).

Despite these developments, most existing studies focus on classroom-based learning or on highly technical applications within specific disciplines. Research that examines the actual use of AI by student interns during OJT remains limited. This gap is particularly evident in the Philippine context, where few studies have explored how students adopt and apply AI tools in workplace-based learning environments. Prior research also specified that training and exposure are important factors in encouraging adoption of new technologies in education (Balilo et al., 2023). To address this gap, the present study investigates how student interns in Philippine higher education engage with AI tools during their OJT. It uses a descriptive-exploratory research design. Specifically, it aims to: (1) describe the types of AI tools used during OJT; (2) determine the frequency and patterns of AI tool usage; (3) identify the main purposes of AI tool use based on task categories; and (4) examine students' perceptions of the effects of AI on their confidence, ethical choices, and supervisory support.

# METHOD

## Research Design, Respondents, and Sampling

This descriptive-exploratory study examined the purposes and perceived effects of intelligent tool usage among student interns in Philippine higher education. The study focused on undergraduate students currently engaged in or recently completing their OJT, immersion, or practicum (subsequently referred as OJT). A survey design was used to collect data on tool usage, frequency, purpose, and perception. Participants were selected through purposive sampling, with inclusion limited to those who had experience using intelligent tools during their OJT. Data were gathered through an online self-administered questionnaire.

## Instrument

The instrument used in this study was developed based on a thorough literature review and operationalized into four constructs, each measured by multiple items. These constructs reflect teachers' attitudes and behaviors toward AI detection tools, including Perceived Effectiveness and Fairness (PEF), Concerns and Peer Influence (CPI), Social Norms and Institutional Influence (SNI), and Trust and Decision-Making (TDM). Each construct assesses different dimensions of teachers' interactions with AI tools, as outlined in Table 1. The instrument underwent validation by two educational technology professors, and reliability testing showed that Cronbach's alpha values for all constructs exceeded the required threshold of .70 which confirmed its internal consistency and reliability. This validated tool provides a comprehensive framework for investigating teachers' perspectives on AI detection tools in academic settings.

## Instrument

The main instrument used in this study was a structured questionnaire composed of five sections. The first section gathered background information, including sex, age, academic program, mode of OJT (on-site, online, or hybrid), personal device usage for OJT tasks, assigned duties, and OJT location. The second section focused on tool usage behavior. It asked respondents to indicate the intelligent tools they used, how often they used these tools during their OJT, and whether they had a paid subscription. The third section measured task-based tool usage which was rated on a 5-point Likert scale, where 1 indicated "Never" and 5 indicated "Very Often.". It consisted items that captured usage during and outside OJT hours. Tasks included but not limited to writing reports, improving grammar, creating presentations, drafting emails, understanding instructions, searching for examples, and receiving coding or technical assistance. The fourth section examined students' perceptions of the effects of using intelligent tools. It included items on confidence, adequacy of skills, perceived usefulness, ethical concerns, and the support received from supervisors. The final section focused on guidance and attitudes toward responsible use. It asked whether students avoided using intelligent tools when the task required independent effort and whether they used such tools only when necessary. The questionnaire was pre-tested with 30 student interns to check the clarity of the items. Based on the feedback, adjustments were made to improve the instrument. Cronbach's alpha was computed for the perception items, and all constructs demonstrated acceptable reliability (i.e., >0.70).

## Data Collection Procedure

Data were collected from May to July 2025 through an online questionnaire. Prior to distribution, permission was secured from participating institutions, starting at the university level, then through their colleges, and finally from OJT or OJT supervisors. The survey was shared with student interns who were currently in or had recently completed their OJT. Participation was voluntary. Respondents were informed about the study's purpose prior the start of the survey and assured of confidentiality. Only those who reported using intelligent tools during their OJT were included in the analysis.

## Statistical Treatment

The study used descriptive statistics, including frequency counts, percentages, and means, to describe the respondents' demographic profile, OJT-related characteristics, and tool usage behavior. To determine the underlying dimensions of the purposes and perceived effects of intelligent tool usage, the study applied factor analysis. It used principal component extraction and Varimax rotation. The Kaiser-Meyer-Olkin (KMO) measure and Bartlett's Test of Sphericity were computed to check whether the data were suitable for factor analysis. Only items with factor loadings of 0.50 or higher were kept. Cronbach's alpha was also computed to test the internal consistency of the grouped items. A value of 0.70 or higher was considered acceptable. All statistical procedures were performed using IBM SPSS v.25.

# RESULTS AND DISCUSSION

## Profile of Respondents

A total of 384 student interns participated in the study. The distribution was nearly balanced, with 197 males and 187 females. Most participants were 22 years old, followed by those aged 23 and 21. The respondents came from a wide range of academic programs across public and private institutions. A large majority (90.7%) completed their OJT on-site, while others did so through hybrid or online arrangements. In terms of device use, 48.7% consistently used their own laptops or phones, 37.1% used them depending on the task, and 14.2% did not use personal devices at all, a situation that may potentially limit opportunities for AI-assisted tasks, as noted by (Balahadia, 2022; Fabito et al., 2021) in the context of online learning challenges. The most common OJT placements were offices (46.4%), schools (22.0%), and IT companies (9.0%). Other placements included hospitals, hotels, government agencies, and utility providers. Students frequently reported performing multiple tasks such as assisting staff or teachers, encoding, data entry, front desk duties, report writing, programming, networking, and design.

## AI Tool Usage Pattern

All respondents provided information regarding their use of AI tools during their OJT (Table 1). Nearly half (48%, n = 296) reported using ChatGPT, making it the most widely used tool. Other commonly used tools included Quillbot (16.9%, n = 104), Canva AI (15.2%, n = 94), and Grammarly (10.5%, n = 65). Smaller percentages used Gemini (3.2%, n = 20), Slidesgo AI (0.6%, n = 4), and Perplexity (0.3%, n = 2). A few respondents (1.5%, n = 9) indicated no use of AI tools or gave unclear responses. Most participants (90.1%, n = 346) relied on free versions of these tools, while 9.9% (n = 38) had subscriptions to premium services. In terms of frequency, 45.3% (n = 174) used AI tools one to two times per week, 11.7% (n = 45) used them three to four times per week, and 7.0% (n = 27) used them daily. A small number (1.6%, n = 6) used AI tools five to six times weekly. Conversely, 34.4% (n = 132) reported not using any AI tools during their OJT.

**Table 1.**
*Frequency and percentage of AI tool usage among student interns*

| AI Tool | Frequency (n) | Percentage (%) |
|---|---|---|
| ChatGPT | 296 | 48 |
| Quillbot | 104 | 16.9 |
| Canva AI | 94 | 15.2 |
| Grammarly | 65 | 10.5 |
| Gemini | 20 | 3.2 |
| Slidesgo AI | 4 | 0.6 |
| Perplexity | 2 | 0.3 |
| Others | 13 | 3.4 |

## Task-based AI Tool Usage

Student interns reported different frequencies of AI tool use depending on the type of task and whether it was done during or outside OJT hours (Table 2). During their OJT, they most often used AI tools to fix grammar or sentence structure (M = 2.80, SD = 1.31), search for sample answers or examples (M = 2.82, SD = 1.30), and understand work instructions or assigned tasks (M = 2.79, SD = 1.33). AI tools were used less frequently for coding or technical assistance (M = 2.70, SD = 1.36), writing drafts of emails or letters (M = 2.63, SD = 1.31), creating slides, posters, or visual materials (M = 2.59, SD = 1.35), and polishing or writing reports (M = 2.56, SD = 1.28). A similar pattern was observed outside OJT hours. Grammar correction (M = 2.91, SD = 1.27), searching for examples (M = 2.84, SD = 1.25), and understanding instructions (M = 2.84, SD = 1.28) remained the most frequent uses. Meanwhile, tasks such as designing slides (M = 2.73, SD = 1.32), drafting communication (M = 2.69, SD = 1.27), seeking coding help (M = 2.68, SD = 1.29), and writing reports (M = 2.72, SD = 1.28) received lower levels of AI support.

These results show that student interns primarily relied on AI tools for language-related and information-seeking purposes, which mirrors how these tools are often used in academic settings (Ozfidan, 2024; Vieriu & Petrea, 2025). Improving grammar, understanding instructions, and locating relevant examples appear to be the most valued functions, as these are essential in completing general work assignments efficiently (Black & Tomlinson, 2025; Malik et al., 2023). In contrast, technical and creative tasks such as coding, design, and report writing received less support from AI tools. This may suggest that students either feel more confident performing those tasks without assistance or view AI tools as less helpful in those areas (Marrone et al., 2022; Verdenhofa et al., 2024). The consistency in usage patterns across OJT and non-OJT settings further suggests that students carry over their AI usage habits regardless of context.

**Table 2.**
*Means and standard deviations for AI tool usage*

| Task | During | | Outside | |
|---|---|---|---|---|
| | **Mean** | **SD** | **Mean** | **SD** |
| Helping write or polish reports | 2.56 | 1.28 | 2.72 | 1.28 |
| Fixing grammar or sentence structure | 2.8 | 1.31 | 2.91 | 1.27 |
| Understanding work instructions/tasks | 2.79 | 1.33 | 2.84 | 1.28 |
| Making drafts of emails/letters | 2.63 | 1.31 | 2.69 | 1.27 |
| Creating slides, posters, or designs | 2.59 | 1.35 | 2.73 | 1.32 |
| Searching for sample answers/examples | 2.82 | 1.3 | 2.84 | 1.25 |
| Coding or technical help (if any) | 2.7 | 1.36 | 2.68 | 1.29 |

## Purposes of Using Intelligent Tools

Exploratory factor analysis on task-based AI tool usage items identified four factors that explained a significant portion of the variance. The first factor, *General Productivity and Report Writing*, contained items like helping write or polish reports (loading = 0.67), fixing grammar (0.72), and understanding instructions (0.55). The second factor, *Communication and Content Drafting*, consisted of tasks such as drafting emails or letters (loading = 0.72), creating slides or posters (0.66), and searching for sample answers (0.64). The third factor, *Technical Assistance and Code Support*, involved coding help and related technical tasks (loading = 0.68). The fourth factor, *Autonomous Task Execution*, included understanding instructions (loading = 0.46) and completing tasks independently (loading = 0.60).

These results suggest that student interns used AI tools for a range of functional purposes, including improving productivity, enhancing communication, supporting technical work, and promoting independent task completion (Agrawal et al., 2024; Hubballi et al., 2025). This supports earlier studies that students primarily engage AI tools to improve clarity, efficiency, and self-management in completing their assignments (Singh et al., 2025). It also reflects trends identified by (Cordevilla et al., 2024), who noted the emerging role of AI in assisting students with various academic and internship-related outputs (Abas et al., 2025).

## Perceived Effects of AI Tool Usage

Student interns felt confident about AI tools for their OJT tasks (M = 3.81, SD = 1.09). They agreed that AI tools helped them finish tasks more quickly and correctly (M = 3.76, SD = 1.06). They moderately agreed that they avoided using AI tools when the task required independent effort (M = 3.14, SD = 1.31). Most reported that supervisors or OJT sites did not stop them from using AI tools (M = 3.60, SD = 1.37). Interns also indicated that they used AI tools selectively and only when necessary (M = 3.42, SD = 1.24). These indicate that student interns adopted a cautious but confident approach to using AI tools (Fošner, 2024; Garcia, 2023; Stritto et al., 2024; Zeki et al., 2025). They valued the benefits of increased efficiency and accuracy, while also showing discernment in tasks where independent effort was expected (Bringula, 2024; Darwin et al., 2024; Hernandez et al., 2025b; Zeki et al., 2025), echoing concerns raised by (Acut et al., 2025) about unverifiable or non-existent sources and the need to verify AI outputs.

These patterns suggest that AI tools improved efficiency but sometimes limited skill development (Gerlich, 2025). Studies show that AI can help people finish tasks faster but may lower accuracy in some cases (Acqua et al., 2023; Zhai et al., 2024). Other research suggests that supervised and reflective use of AI, where learners edit or adapt the output, may strengthen learning outcomes (Bringula, 2024; Wei, 2023; Xu, 2024). The presence of a supervisor or clear guidelines may influence how students (including interns) use AI (Bringula, 2023; Jin et al., 2025). Without these, they may rely too much on AI, which can reduce opportunities to practice and improve skills (Jin et al., 2025). In contrast, those who use AI to check or refine their work while continuing to engage with the task may further develop their abilities (Vieriu & Petrea, 2025).

## CONCLUSION AND RECOMMENDATIONS

This study examined the use of intelligent tools by student interns in Philippine higher education during their OJT. ChatGPT was the most frequently used AI tool, followed by Quillbot, Canva AI, and Grammarly. Most students used AI one to two times per week, mainly for productivity tasks such as report writing, grammar checking, and clarifying work instructions. AI also supported communication tasks like drafting emails and creating presentations, as well as technical work such as coding assistance. Students in this study reported moderate confidence in using AI tools and showed ethical awareness by avoiding AI when independent effort was required. Supervisors generally allowed AI use, and students applied it selectively depending on the task. However, variations in confidence levels, ethical judgment, and patterns of use suggest the need for structured AI literacy training in OJT programs, along with clear workplace guidelines that help interns maximize benefits while maintaining ethical standards.

While these results provide useful insights, they also highlight areas that require deeper examination. Future research should go beyond describing usage patterns and examine how AI tools affect learning, skill development, and workplace readiness. This study offers an initial look at how interns use AI in the Philippine context, and the study point to the need for further investigation. Future work can use statistical methods to explore links between AI use and performance outcomes, and compare patterns across different industries and workplace settings in the country. Qualitative studies with interns and supervisors can also provide more insight into how decisions about AI use are made. These steps can help build a clearer understanding of AI's role in workplace learning and support more targeted policies and practices for OJT programs.

## PRACTICAL IMPLICATIONS

AI tools such as ChatGPT are becoming part of the skill-building process for student interns in the Philippines, particularly in developing communication, problem-solving, and technical abilities. By taking on routine tasks, these tools allow interns to focus on more complex duties and improve productivity during OJT. In local contexts, interns in provincial government offices may use AI to draft reports, while those in hospitality settings may use it to prepare guest itineraries. Higher education institutions and OJT sites should integrate AI literacy and onboarding programs that reflect workplace realities in sectors like hospitality, tourism, and public service. Clear policies are needed to guide ethical and responsible AI use while encouraging independent thinking. Equitable access to AI resources is also essential, especially for interns in rural workplaces with limited internet access.

Educators should ensure OJT orientations include training on how AI can complement rather than replace work. For example, marketing interns can learn to use AI for campaign outlines but must adapt them using client data from local businesses. Supervisors should set task-specific AI guidelines, such as allowing AI to help in drafting reports but requiring interns to add context-specific details based on their workplace. Institutions should require interns to log AI-assisted work so faculty can review its impact on skill development. In tourism offices, reviewing AI-generated itineraries can ensure they meet client needs and reflect local culture. These steps can help align AI use in OJT with both industry expectations and the goal of building long-term professional competence.

# REFERENCES

Abas, A., Ibrahim, M. M. B., & Azmi, N. H. (2025). Data-Driven Framework for Optimizing Internship Efficiency and Addressing Skills Mismatch in Malaysian Higher Education. *22nd International Learning and Technology Conference: Human-Machine Dynamics Fueling a Sustainable Future, L and T 2025*, 280–285. https://doi.org/10.1109/LT64002.2025.10940990

Acqua, F. D., Lifshitz-assaf, H., Kellogg, K. C., Krayer, L., Acqua, F. D., Krayer, L., & Mollick, E. (2023). *Navigating the Jagged Technological Frontier: Field Experimental Evidence of the Effects of AI on Knowledge Worker Productivity and Quality*. https://www.hbs.edu/faculty/Pages/item.aspx?num=64700

Acut, D. P., Malabago, N. K., Malicoban, E. V, Galamiton, N. S., & Garcia, M. B. (2025). "ChatGPT 4.0 Ghosted Us While Conducting Literature Search:" Modeling the Chatbot's Generated Non-Existent References Using Regression Analysis. *Internet Reference Services Quarterly*, *29*(1), 27–54. https://doi.org/10.1080/10875301.2024.2426793

Agrawal, L., Lanjewar, P., Deshpande, S., Jawarkar, P., Gaur, V., & Dive, A. (2024). The Impact of AI on Communication Skills Training Opportunities and Challenges. *Nanotechnology Perceptions*, *20*(S7), 1167–1173. https://doi.org/10.62441/nano-ntp.v20iS7.96

Al-Sofi, B. B. M. A. (2024). Artificial intelligence-powered tools and academic writing: to use or not to use ChatGPT. *Saudi Journal of Language Studies*, *4*(3), 145–161. https://doi.org/10.1108/SJLS-06-2024-0029

Balahadia, F. (2022). Challenges of Information Technology Education Student's Online Classes during the Covid-19 Pandemic in Selected Public Colleges and Universities in the Philippines. *International Journal of Computing Sciences Research*, *6*, 877–892. https://doi.org/10.25147/ijcsr.2017.001.1.79

Balilo, B. B., Rodriguez, R. A., & Guerrero, J. J. G. (2023). Insights and Trends of Capstone Project of CS / IT Department of Bicol University from 2011 to 2019. *International Journal of Computing Sciences Research*, *7*(273), 1255–1272. https://doi.org/10.25147/ijcsr.2017.001.1.99

Black, R. W., & Tomlinson, B. (2025). University students describe how they adopt AI for writing and research in a general education course. *Scientific Reports*, *15*(1), 8799. https://doi.org/10.1038/s41598-025-92937-2

Bringula, R. (2023). What do academics have to say about ChatGPT? A text mining analytics on the discussions regarding ChatGPT on research writing. *AI and Ethics*. https://doi.org/10.1007/s43681-023-00354-w

Bringula, R. (2024). ChatGPT in a programming course: benefits and limitations . In *Frontiers in Education* (Vol. 9).

Castillo-Núñez, N., Mejía-Díaz, V., Varas-Reyes, J., & Avello, D. (2024). Therapeutic communication skills in Occupational Therapy students during professional practice: perceptions and opinions from the faculty. *Revista Medica Clinica Las Condes*, *35*(5–6), 473–483. https://doi.org/10.1016/j.rmclc.2024.10.003

Catacutan, K. J. A., & Tuliao, A. S. (2020). On-the-Job Training Program Evaluation of Business Administration and Accountancy Departments of University of Saint Louis. *Universal Journal of Educational Research*, *8*(1), 143–150. https://doi.org/10.13189/ujer.2020.080117

Cordevilla, R. P., Banot, V. L., & Velez, J. M. M. (2024). Generative AI: perspectives of teaching interns. *Science International-(Lahore)*, *36*(6), 665–675. https://sci-int.com/pdf/638700548309490016.Cordevilla- FINAL Generative AI Perspectives of Teaching Interns (1) (1) (1).edited.pdf

Darwin, Rusdin, D., Mukminatien, N., Suryati, N., Laksmi, E. D., & Marzuki. (2024). Critical thinking in the AI era: An exploration of EFL students' perceptions, benefits, and limitations. *Cogent Education*, *11*(1), 2290342. https://doi.org/10.1080/2331186X.2023.2290342

Debliquy, N., Coppe, T., Deschepper, C., & Colognesi, S. (2025). Experiencing difficulty in internships as a catalyst for improvement in pre-service teachers' reflective writing skills. *Teacher Development*. https://doi.org/10.1080/13664530.2025.2489413

Fabito, B. S., Trillanes, A. O., & Sarmiento, J. (2021). Barriers and Challenges of Computing Students in an Online Learning Environment: Insights from One Private University in the Philippines. *International*

*Journal of Computing Sciences Research*, *5*(1), 441–458. https://doi.org/10.25147/ijcsr.2017.001.1.51

Fošner, A. (2024). University Students' Attitudes and Perceptions towards AI Tools: Implications for Sustainable Educational Practices. In *Sustainability* (Vol. 16, Issue 19). https://doi.org/10.3390/su16198668

Garcia, M. B. (n.d.). ChatGPT as an Academic Writing Tool: Factors Influencing Researchers' Intention to Write Manuscripts Using Generative Artificial Intelligence. *International Journal of Human–Computer Interaction*, 1–15. https://doi.org/10.1080/10447318.2025.2499158

Garcia, M. B. (2023). ChatGPT as a Virtual Dietitian: Exploring Its Potential as a Tool for Improving Nutrition Knowledge. In *Applied System Innovation* (Vol. 6, Issue 5). https://doi.org/10.3390/asi6050096

Gerlich, M. (2025). AI Tools in Society: Impacts on Cognitive Offloading and the Future of Critical Thinking. In *Societies* (Vol. 15, Issue 1). https://doi.org/10.3390/soc15010006

Hernandez, H. E., Manalese, R. P., Dianelo, R. F. B., Yambao, J. A., Gamboa, A. B., Feliciano, L. D., David, M. H. M., Pampo, F. R., & Miranda, J. P. P. (2025a). Dependency on Meta AI Chatbot in Messenger Among STEM and Non-STEM Students in Higher Education. *International Journal of Computing Sciences Research; Vol 9 (2025): Volume 9*. file://stepacademic.net/ijcsr/article/view/660

Hernandez, H. E., Manalese, R. P., Dianelo, R. F. B., Yambao, J. A., Gamboa, A. B., Feliciano, L. D., David, M. H. M., Pampo, F. R., & Miranda, J. P. P. (2025b). Dependency on Meta AI Chatbot in Messenger Among STEM and Non-STEM Students in Higher Education. *International Journal of Computing Sciences Research*, *9*, 3674–3690. https://doi.org/10.25147/ijcsr.2017.001.1.242

Hingpit, S. P. G. (2024). The On-The-Job Training Experiences of Business Administration Students of Philippine Electronics and Communication Institute of Technology Butuan City Authors Sheigfred Paulo G. Hing. *JPAIR Multidisciplinary Research*, *57*(1), 83–112. https://doi.org/https://doi.org/10.7719/jpair.v57i1.894

Hubballi, R. B., Selvakumar, P., Seenivasan, R., Guru Basava Aradhya, S., Dinesh, N., & Seelam, P. K. (2025). Overview of current AI technologies in education. In *Transformative AI Practices for Personalized Learning Strategies* (pp. 1–25). https://doi.org/10.4018/979-8-3693-8744-3.ch001

Jin, Y., Yan, L., Echeverria, V., Gašević, D., & Martinez-Maldonado, R. (2025). Generative AI in higher education: A global perspective of institutional adoption policies and guidelines. *Computers and Education: Artificial Intelligence*, *8*, 100348. https://doi.org/https://doi.org/10.1016/j.caeai.2024.100348

Keawtavon, T., Suamuang, W., & Chomsuwan, K. (2023). Professional Competency Development Using Self-Directed Learning via on the Job Training. *8th International STEM Education Conference, ISTEM-Ed 2023 - Proceedings*. https://doi.org/10.1109/iSTEM-Ed59413.2023.10305790

Loc, T. B., Thanh Ngan, N. T., Linh, L. D., & Luan, N. T. (2025). On-The-Job Training: Enhance Experience and Increase Success for the Student Community in Vietnam. *SAGE Open*, *15*(1). https://doi.org/10.1177/21582440241311160

Malik, A. R., Pratiwi, Y., Andajani, K., Numertayasa, I. W., Suharti, S., Darwis, A., & Marzuki. (2023). Exploring Artificial Intelligence in Academic Essay: Higher Education Student's Perspective. *International Journal of Educational Research Open*, *5*, 100296. https://doi.org/https://doi.org/10.1016/j.ijedro.2023.100296

Marrone, R., Taddeo, V., & Hill, G. (2022). Creativity and Artificial Intelligence—A Student Perspective. In *Journal of Intelligence* (Vol. 10, Issue 3). https://doi.org/10.3390/jintelligence10030065

Marzuki, , Widiati, U., Rusdin, D., Darwin, , & Indrawati, I. (2023). The impact of AI writing tools on the content and organization of students' writing: EFL teachers' perspective. *Cogent Education*, *10*(2), 2236469. https://doi.org/10.1080/2331186X.2023.2236469

Öncü, S., Torun, F., & Ülkü, H. H. (2025). AI-powered standardised patients: evaluating ChatGPT-4o's impact on clinical case management in intern physicians. *BMC Medical Education*, *25*(1). https://doi.org/10.1186/s12909-025-06877-6

Ozfidan, B. (2024). The use of AI tools in English academic writing by Saudi undergraduates. *Contemporary Educational Technology*, *16*(4), 1–13.

Salutina, T. Y., Platunina, G. P., & Frank, I. A. (2024). The Role of Artificial Intelligence in Improving the Effectiveness of Professional Training of Students of Electrical Engineering and Electronic Engineering. *2024 Intelligent Technologies and Electronic Devices in Vehicle and Road Transport*

*Complex, TIRVED 2024 - Conference Proceedings*. https://doi.org/10.1109/TIRVED63561.2024.10769921

Sanahuja Vélez, G., Ribes Giner, G., & Moya Clemente, I. (2017). Intrapreneuring Within a Higher Education Institution: Introducing Virtual Business Internships. In *Innovation, Technology and Knowledge Management* (pp. 259–266). https://doi.org/10.1007/978-3-319-47949-1_18

Schartel Dunn, S. G., & Lane, P. L. (2019). Do Interns Know What They Think They Know? Assessing Business Communication Skills in Interns and Recent Graduates. *Business and Professional Communication Quarterly*, *82*(2), 202–213. https://doi.org/10.1177/2329490619826258

Singh, A. K., Kiriti, M. K., Singh, H., & Shrivastava, A. (2025). Education AI: exploring the impact of artificial intelligence on education in the digital age. *International Journal of System Assurance Engineering and Management*, *16*(4), 1424–1437. https://doi.org/10.1007/s13198-025-02755-y

Stritto, M. E. Dello, Underhill, G. R., & Aguiar, N. R. (2024). *Online Students' Perceptions of Generative AI*. https://ecampus.oregonstate.edu/research/wp-content/uploads/Online-Students-Perceptions-of-AI-Report.pdf

Verdenhofa, O., Kinderis, R., & Berjozkina, G. (2024). AI as a creative partner: how artificial intelligence impacts student creativity and innovation: case study of students from Latvia, Ukraine and Spain. *Baltic Journal of Economic Studies*, *10*(5), 111–119. https://doi.org/https://doi.org/10.30525/2256-0742/2024-10-5-111-119

Vieriu, A. M., & Petrea, G. (2025). The Impact of Artificial Intelligence (AI) on Students' Academic Development. In *Education Sciences* (Vol. 15, Issue 3). https://doi.org/10.3390/educsci15030343

Wang, Z., Zhu, Y., Zhan, X., Wang, T., Tang, X., Li, L., Su, T., Zhou, H., Liu, L., Chen, L., Pang, X., Peng, J., Wang, Y., & Yang, L. (2024). Problem-solving ability and future time perspective among the Chinese nursing interns: The mediating role of future work self. *PLoS ONE*, *19*(8 August). https://doi.org/10.1371/journal.pone.0308669

Weaver, C., Caputo, N. J., Yost, C., & Hauert, S. A. (2025). Using an AI-Nurse to Help Teach Graduating Medical Students to Take Overnight Call. *Medical Science Educator*, *35*(2), 615–617. https://doi.org/10.1007/s40670-025-02335-6

Wei, L. (2023). Artificial intelligence in language instruction: impact on English learning achievement, L2 motivation, and self-regulated learning. *Frontiers in Psychology*, *14*.

Xu, Z. (2024). AI in education: Enhancing learning experiences and student outcomes. *Proceedings of the 4th International Conference on Signal Processing and Machine Learning*, 104–111. https://doi.org/10.54254/2755-2721/51/20241187

Yamauchi, F., Kim, T., Lee, K. W., & Tiongco, M. (2023). Can German vocational training combat skill shortages in developing countries? Evidence from dual training system in the Philippines. *Review of Development Economics*, *27*(4), 2470–2488. https://doi.org/10.1111/rode.13047

Zeki, A. M., Tarshany, Y., Wasiq, S., Al-Taei, M. H. A., & Alhazmi, A. K. (2025). Exploring University Students' Utilization of AI Tools in Academic Assignments: Benefits, Challenges, and Ethical Considerations. *2025 International Conference for Artificial Intelligence, Applications, Innovation and Ethics (AI2E)*, 1–6. https://doi.org/10.1109/AI2E64943.2025.10983650

Zhai, C., Wibowo, S., & Li, L. D. (2024). The effects of over-reliance on AI dialogue systems on students' cognitive abilities: a systematic review. *Smart Learning Environments*, *11*(1), 28. https://doi.org/10.1186/s40561-024-00316-7